\documentclass[useAMS,usenatbib]{mn2e}
\usepackage{epsfig}

\title[ISO-Oph-50]{The young low-mass star ISO-Oph-50: Extreme variability induced by a clumpy, evolving circumstellar disk}

\author[Scholz, Mu\v{z}i\'c, Geers]{Alexander Scholz$^{1}$\thanks{E-mail: as110@st-andrews.ac.uk}, 
Koraljka Mu\v{z}i\'c$^{2}$, Vincent Geers$^{3}$\\
$^{1}$School of Physics and Astronomy, University of St. Andrews, The North Haugh, St. Andrews, Fife, KY16 9SS, United Kingdom\\
$^{2}$European Southern Observatory, Alonso de C\'ordova 3107, Casilla 19, Santiago 19001, Chile\\
$^{3}$UK Astronomy Technology Centre, Royal Observatory Edinburgh, Blackford Hill, Edinburgh, EH9 3HJ, United Kingdom\\ 
}

\begin{document}

\date{Accepted. Received.}

\pagerange{\pageref{firstpage}--\pageref{lastpage}} \pubyear{2002}

\maketitle

\label{firstpage}

\begin{abstract}
ISO-Oph-50 is a young low-mass object in the $\sim 1$\,Myr old Ophiuchus star forming region undergoing dramatic changes in its
optical/near/mid-infrared brightness by 2-4\,mag. We present new multi-band photometry and near-infrared spectra, combined
with a synopsis of the existing literature data. Based on the spectroscopy, the source is confirmed as a mid
M dwarf, with evidence for ongoing accretion. The near-infrared lightcurves show large-scale variations, with 2-4\,mag
amplitude in the bands IJHK, with the object generally being bluer when faint. Near its brightest state, the object shows 
colour changes consistent with variable extinction of $\Delta A_V \sim 7$\,mag. High-cadence monitoring at 3.6$\,\mu m$ reveals 
quasi-periodic variations with a typical timescale of 1-2 weeks. The best explanation for these characteristics is 
a low-mass star seen through circumstellar matter, whose complex variability is caused by changing inhomogeneities in the inner
parts of the disk. When faint, the direct stellar emission is blocked; the near-infrared radiation is dominated by 
scattered light. When bright, the emission is consistent with a photosphere strongly reddened by circumstellar dust. 
Based on the available constraints, the inhomogeneities have to be located at or beyond $\sim 0.1$\,AU distance from 
the star. If this scenario turns out to be correct, a major portion of the inner disk has to be clumpy, structured, and/or 
in turmoil. In its observational characteristics, this object resembles other types of YSOs with variability caused 
in the inner disk. Compared to other objects, however, ISO-Oph-50 is clearly an extreme case, given the large amplitude of 
the brightness and colour changes combined with the erratic behaviour. ISO-Oph-50 has been near its brightest state since 2013; 
further monitoring is highly encouraged.
\end{abstract}


\section{Introduction}
\label{s0}

Variability has long been known as a characteristic observational signature of young stars. Most common are lightcurves 
with periods of days to weeks due to spots on the stellar surface, a fact that has been used for almost three 
decades to infer the rotation periods of young stars \citep{1989A&A...211...99B,2007prpl.conf..297H}.  
Apart from these regular changes, a fraction of young stellar objects (YSOs) shows evidence for large-scale, partly 
irregular variations \citep{1994AJ....108.1906H}. This was originally one of the criteria for the definition of the class 
of 'T Tauri stars' \citep{1945ApJ...102..168J}. Some of the irregular variability follows patterns that are well-explained by 
astrophysical phenomena, including hot accretion-induced spots \citep{1989AJ.....97..483V,1995A&A...299...89B}, magnetic flares 
\citep{1983ApJ...267..191R,1995AJ....110..336W}, accretion bursts 
\citep[the FU Ori and EX Lupi events;][]{1989ESOC...33..233H,1996ARA&A..34..207H}, and obscurations by circumstellar 
material. The latter includes the UX Ori phenomena for Herbig Ae/Be stars \citep{1991Ap&SS.186..283G}, but also other types
\citep{2001ApJ...554L.201H,2012A&A...544A.112R}. Variable extinction may be the reason for the 
extreme fluctuations observed in the prototype, T Tauri \citep{2001AJ....122..413B}. Often young stars are affected
by several of these processes. For a comprehensive discussion of the various factors that contribute to YSO variability 
and a discussion of possible models, see \citet{2001AJ....121.3160C,2009MNRAS.398..873S,2013ApJ...773..145W}.

Although extreme variables have been defining prototypes for young stellar objects, their overall fraction is 
low. In the near-infrared, only 2-3\% of an unbiased sample of YSOs show variations larger than 0.5\,mag over timescales up
to several years \citep{2012MNRAS.420.1495S}. In the optical, that fraction rises to $\sim 10$\% (Rigon et al., in prep.). 
The subset of YSOs variable on timescales of years is only 10-20\% in the mid-infrared \citep{2014AJ....148...92R}.
The fraction of extremely variable YSOs drops with age; only very few are known older than 
5\,Myr \citep{2012AJ....143...72M,2010A&A...524A...8G}, in line with typical disk lifetimes. Despite their rarity, 
YSOs with strong variations present unique insights into specific processes and phases occuring during the first few million years in a 
star's life (e.g., episodes of strong accretion). In particular, YSOs which are obscured by their disks can provide information about 
the structure and the evolution of circumstellar accretion disks, the properties of the dust grains in the vicinity of the star, 
as well as the initial conditions of planet formation
\citep[e.g.][]{2002A&A...393..259M,2008Natur.452..194H,2013AJ....146..112R}.

So far, most of the detailed studies of extreme T Tauri variability have been carried out in the optical and for 
stars more massive or similar in mass to the Sun. Current studies are driven by large-scale optical variability surveys 
(e.g., Super-WASP, KELT, PTF, Kepler, Gaia). Many deeply embedded and very low mass YSOs 
are not covered by such surveys. On the other hand, given the fact that most stars have masses $<0.5\,M_{\odot}$ and that 
the occurence of extreme variability declines with age, it is interesting to investigate the aforementioned phenomena for 
lower-mass objects and in the embedded phase, to probe how applicable the proposed explanations are to the youngest 
and more typical YSOs. As an example, it was recently found that an embedded very low mass star undergoes an FU Ori-like 
outburst, demonstrating that accretion bursts can occur over a wide range of stellar masses \citep{2011A&A...526L...1C}.

Here we present new observations for the infrared source ISO-Oph-50, located in the $\sim 1$\,Myr $\rho$\,Oph star 
forming region\footnote{2MASS J16263682-2419002, J2000 coordinates: 16:26:36.8, -24:19:00}. The object was identified by 
\citet{2001A&A...372..173B} from ISO observations and tentatively classified as Class III (i.e. without disk) based on the 
infrared SED slope. More recent studies based on Spitzer data put it at an earlier evolutionary state of Class I or II 
\citep{2014AJ....148..122G,2012A&A...539A.151A}. ISO-Oph-50 has been known to show extreme variability in the infrared 
\citep{2008A&A...485..155A,2013MNRAS.430.2910S}. So far, the only available spectrum 
shows that the object is an M-dwarf \citep{2012A&A...539A.151A}. The source has excess emission in the mid-infrared indicating 
the presence of circumstellar dust, but so far there is no evidence for ongoing accretion. In this paper we analyse new 
multi-band photometry and near-infrared spectra and discuss possible explanations for the variability.

\section{Observations}
\label{s1}

\subsection{SMARTS photometry}

We used the 1.3\,m telescope at the Cerro Tololo International Observatory to obtain long-term 
photometry for ISO-Oph-50, in the framework of guaranteed observing time through the SMARTS 
collaboration. In total we observed the source in 20 nights spread over 2 years. The telescope is 
equipped with Andicam, a dual-channel camera which takes optical and near-infrared images simultaneously. 
Our observing scripts first take images in I- and J-band, then R- and K-band (with the exception of the
very first epoch in which we observed in H instead of K). ISO-Oph-50 is not 
visible in the R-band frames, and only barely detected in most I-band images. All observations were
done at low airmass ranging from 1.0 to 1.5.

In the near-infrared bands, the object was observed in a 5 position dither pattern with integration 
times of $5 \times 30$\,sec per band. A standard image reduction was carried out, including sky subtraction, 
flatfielding and aperture photometry. While the J-band images show only our target, the H- and K-band images cover
three more stars with 2MASS photometry. We calibrated the photometry in relation to the
2MASS values for these three stars. The offset between instrumental K-band magnitudes $K_{\mathrm{inst}}$ and 
2MASS K-band magnitudes $K_{\mathrm{2M}}$ remains stable at $K_{\mathrm{inst}} - K_{\mathrm{2M}} = 3.0 \pm 0.1$
and does not significantly depend on the colour of the comparison stars. This offset includes the constant
zeropoint and an extinction term, the variations of $\pm 0.1$ are consistent with a combination of photometric 
errors and the varying airmass. Typical errors of the K-band magnitudes are $\pm 0.05$\,mag, combining the 
calibration uncertainties and the photometric error.

\begin{figure}
\includegraphics[width=9.0cm]{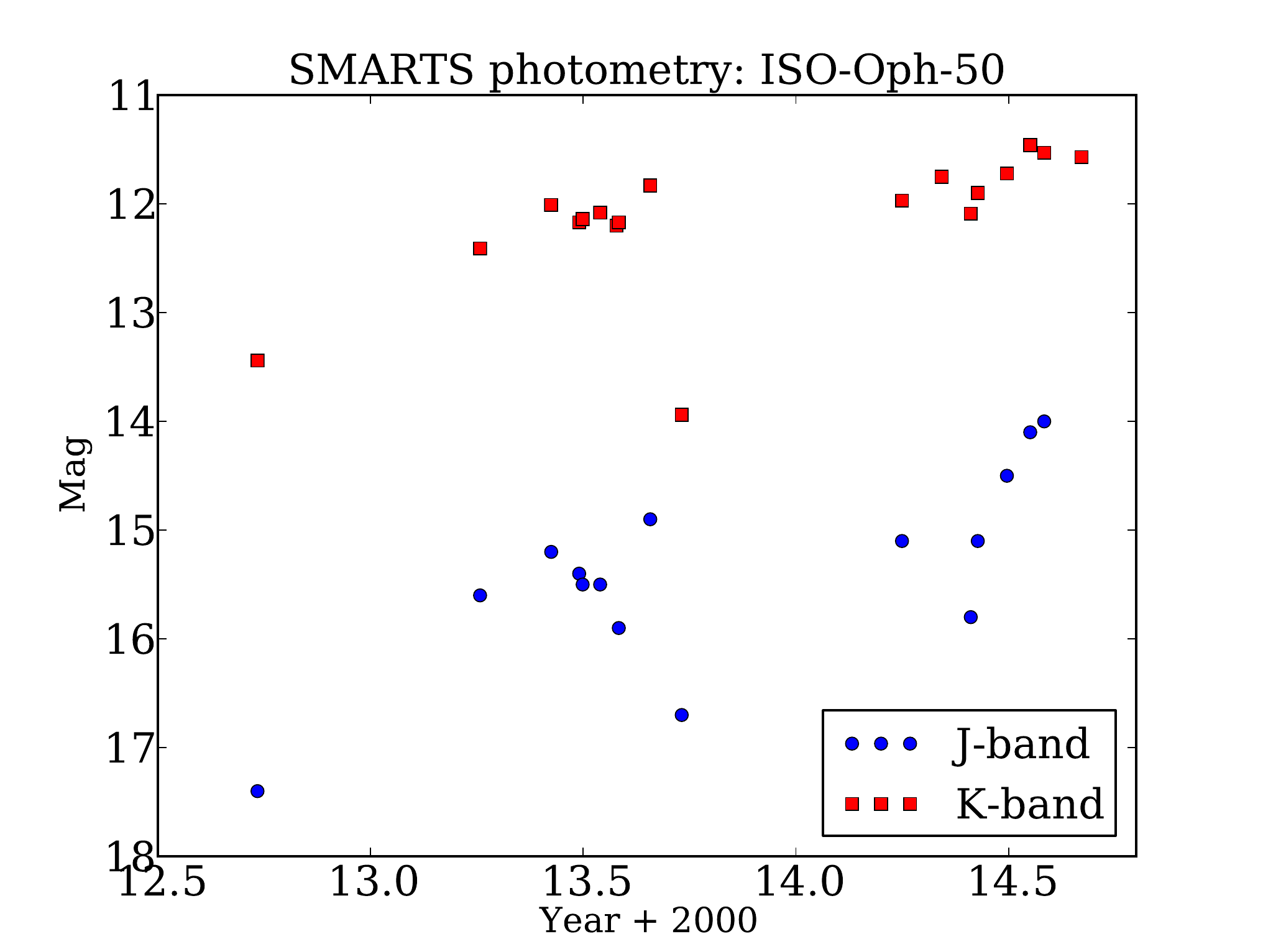}
\caption{\label{lc} SMARTS J/K-band lightcurve for ISO-Oph-50 covering two years. Errors in J and K are $\pm 0.1$ 
and $\pm 0.05$\,mag, respectively.}
\end{figure}

Since we do not have a calibration star in the J-band, we have established the offset between instrumental magnitudes 
and 2MASS magnitudes from observations of other fields, published in \citet{2013MNRAS.430.2910S}. The offset in the
J-band is $J_{\mathrm{inst}} - J_{\mathrm{2M}} \sim 2.5 \pm 0.1$ through seasons 2012 to 2014. Similarly to the
K-band, this value is stable within the quoted errorbars and for the given airmass range. We adopt this offset
for all our observations of ISO-Oph-50. Since we cannot measure the offset for the images themselves, the 
resulting uncertainties may be as large as $\pm 0.1$ (comparable to the variations in the offset). The calibrated 
J- and K-band lightcurves are plotted in Fig. \ref{lc}.

For the optical images, we obtained three images per band at the same position, each with an integration time
of 110\,sec. To increase signal-to-noise, the I-band images were co-added. As calibration star, we used 
the object {\it 2MASS J16264285-2420299} (or ISO-Oph-62), which has a DENIS I-band magnitude of 14.355 and 
does not show signs of significant variability. The uncertainty in the I-band measurements is in the range of
$\pm 0.3$\,mag. This estimate includes the photometric error ($\pm 0.1$\,mag) plus possible calibration systematics.

The full SMARTS photometry is listed in the appendix in table \ref{at1}. This table also includes the first epoch of 
SMARTS photometry in J- and H-band which was already published in \citet{2013MNRAS.430.2910S} for the 12th of August 2012.

\subsection{NTT/SOFI Spectroscopy}

We took two spectra of ISO-Oph-50 using SOFI \citep[Son of ISAAC;][]{1998Msngr..94....7M} at the ESO's New Technology 
Telescope (NTT) during the nights of April 13 and 19, 2014, under the programme number 093.C-0050(A). We used two
low resolution grisms, covering the wavelength range 0.95 - 1.64\,$\mu$m (Blue Grism) and 1.5 - 2.5\,$\mu$m (Red Grism) 
with $1''$ slit, resulting in spectral resolution $R \sim 600$.

Data reduction was performed by combining the SOFI pipeline recipes, our IDL routines, and the IRAF task $apall$. 
The data reduction steps include cross-talk, flat field, and distortion corrections. Pairs of frames at different 
nodding positions were subtracted one from another, shifted, and combined into a final frame. We extracted the spectra
using the IRAF task {\it apall} and wavelength calibrated using the solution obtained by the pipeline. 
The spectra of telluric standard stars were reduced in the same way as the science frames. Telluric standards (spectral 
types B2V and B5V) show several prominent hydrogen lines in absorption, which we removed from the spectra by 
interpolation, prior to division with the black body spectrum at an appropriate effective temperature. This yields
the response function of the spectrograph. Finally, we divided the science spectra by the corresponding response
function.

\begin{figure}
\includegraphics[width=8.6cm]{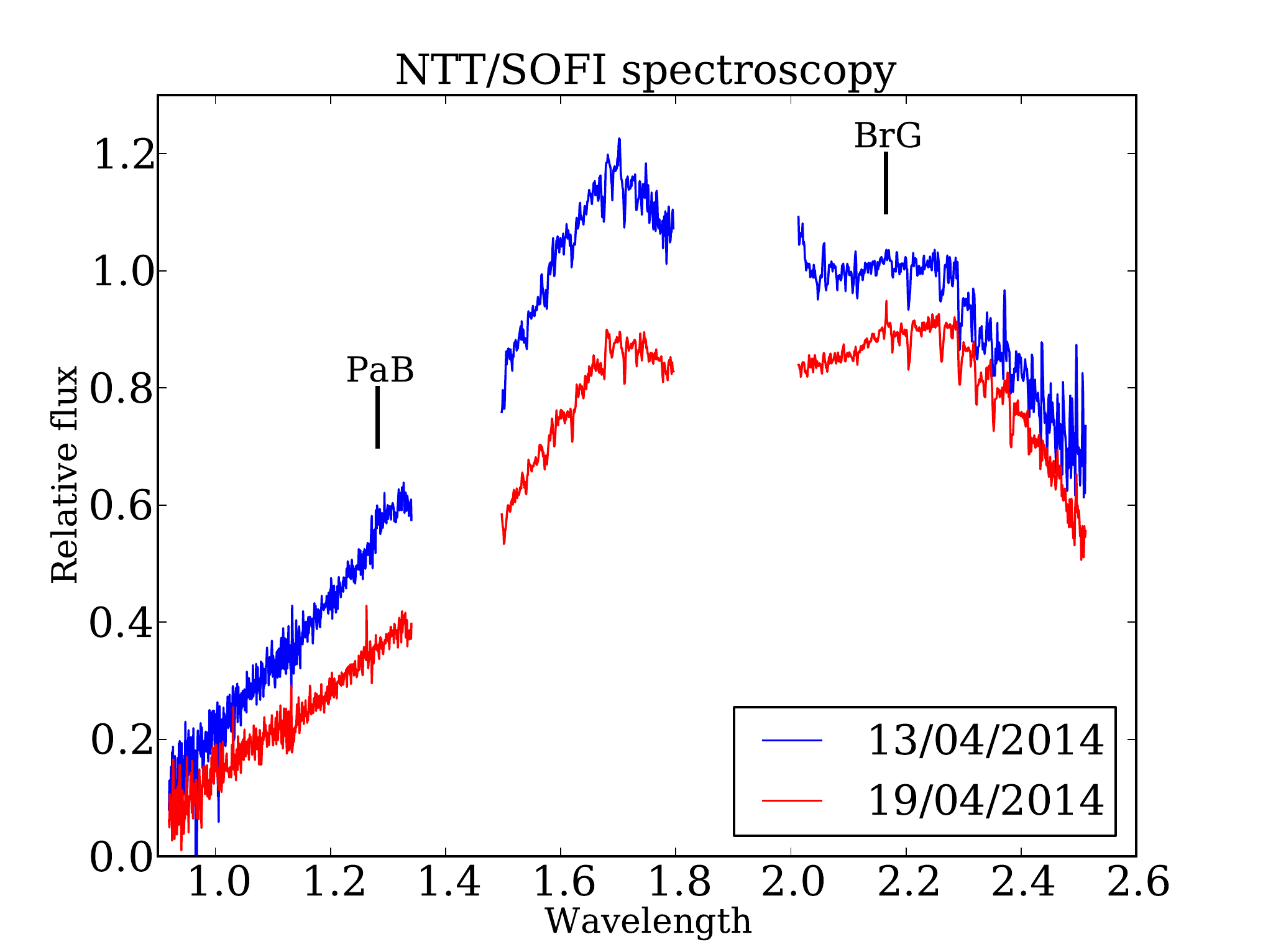}
\caption{\label{spec} NTT/SOFI spectra for ISO-Oph-50 taken on April 13th and 19th 2014. The two spectra are calibrated against
each other using the K-band magnitudes obtained from the acquisition images. The accretion-related lines Paschen\,$\beta$ and 
Bracket\,$\gamma$ are marked.}
\end{figure}

\begin{figure*}
\includegraphics[width=8.6cm]{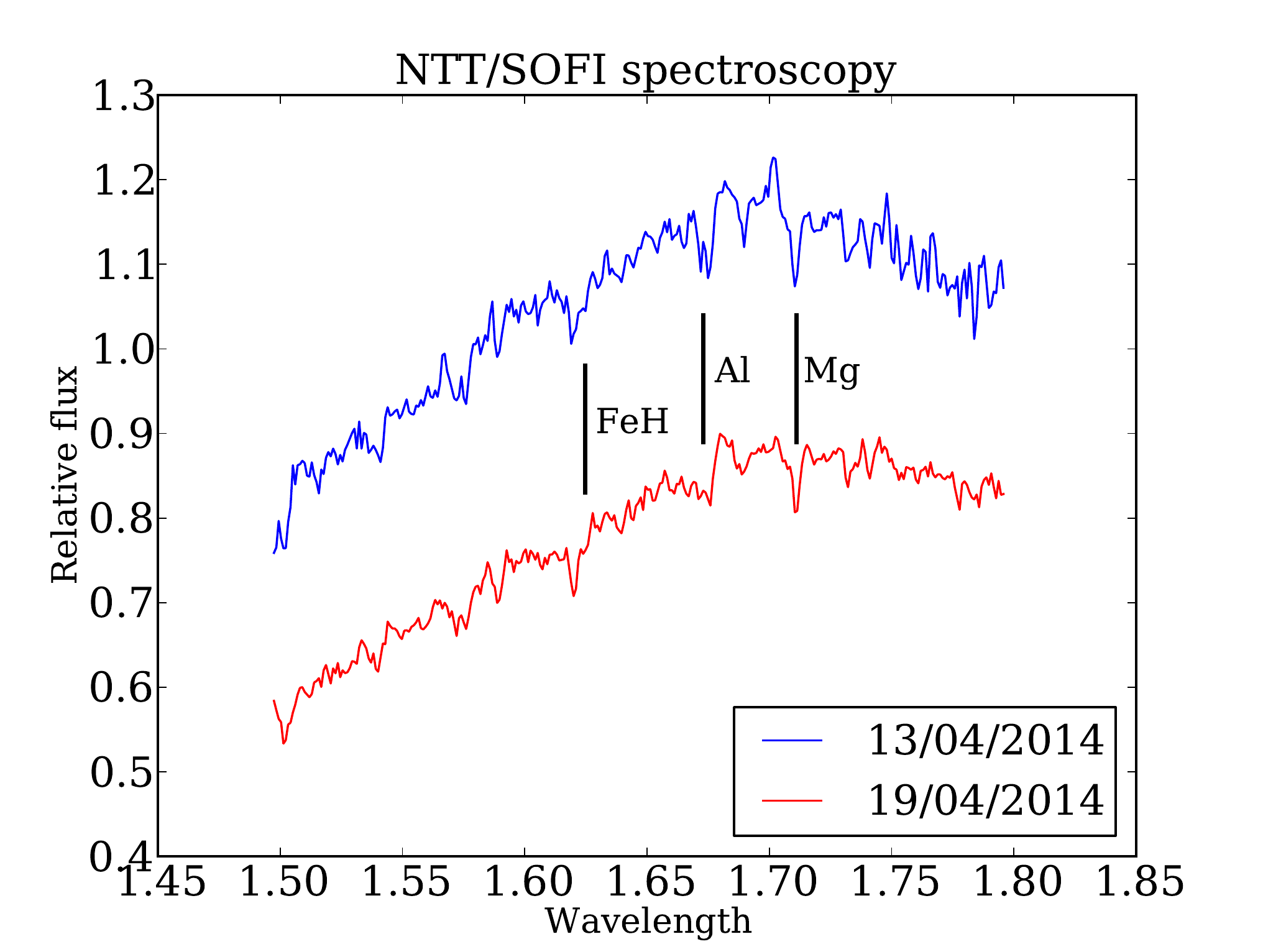}
\includegraphics[width=8.6cm]{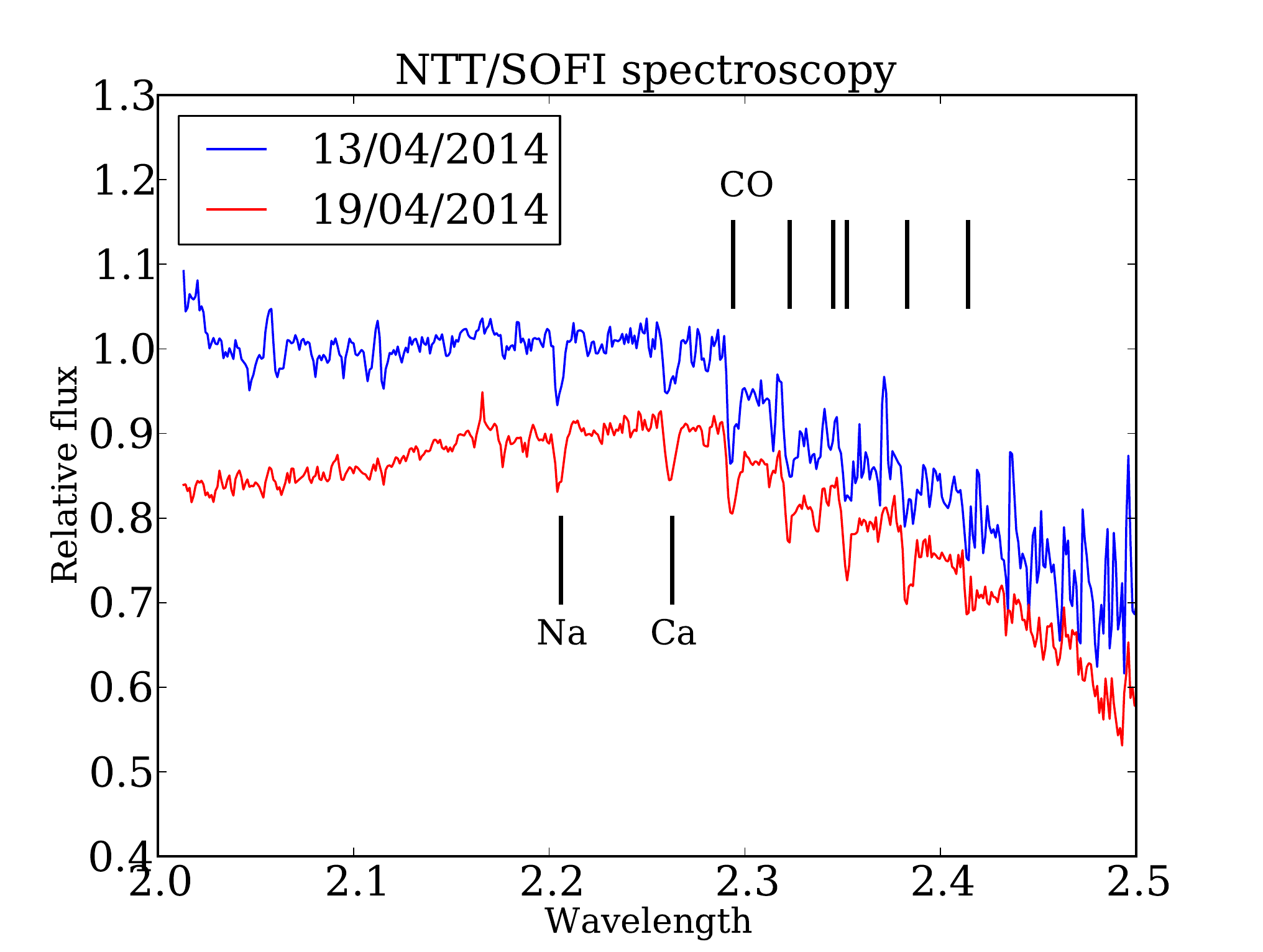}
\caption{\label{spechk} NTT/SOFI spectra for ISO-Oph-50 taken on April 13th and 19th 2014; close-up view of the H- and K-band.
Several photospheric absorption features are marked.}
\end{figure*}

We also reduced the K-band acquisition images and measured magnitudes of 11.85 and 12.00 for April 13 and 19th, 
respectively. These values were derived in comparison with several 2MASS stars in the same field of view. We scaled
the spectra so that the ratio of the integrated K-band fluxes corresponds to the flux ratio as determined from the
photometry. We also scaled each of the two blue spectra so that they match the corresponding red spectrum in the overlapping
wavelength range in the H-band. That way, although the two spectra are not calibrated in an absolute flux scale, they 
can be compared with each other across the entire wavelength regime. The full spectra are shown in Fig. \ref{spec}; 
the H- and K-band are plotted separately in Fig. \ref{spechk}.

\subsection{Mid-infrared photometry}

ISO-Oph-50 has been observed multiple times with the IRAC instrument on-board Spitzer, mostly in the framework of the 
project YSOVAR (PI: J. Stauffer, see \citet{2011ApJ...733...50M,2014AJ....148...92R}). The $\rho$\,Oph survey as part of YSOVAR has
recently been published by \citet{2014AJ....148..122G}. In their paper, ISO-Oph-50 is found to be strongly variable with
a reduced $\chi ^2$ of 821.89, second highest in their sample, confirming the exceptional nature of this source. We 
compiled all available images for ISO-Oph-50 taken 
in the Spitzer channels 1 and 2 (3.6 and 4.5$\,\mu m$) and measured the flux of the source. In total, we obtained 104 epochs 
in IRAC1 and 20 epochs in IRAC2. Fluxes were converted to magnitudes using appropriate aperture corrections and zeropoints 
of 281 and 180\,Jy for channels 1 and 2, respectively. The calibration was verified by comparing with the publicly available 
fluxes from the C2D Legacy project \citep{2009ApJS..181..321E}. Typical uncertainties in the Spitzer fluxes are 
$\pm 0.05$\,mag. We note that the YSOVAR lightcurve (see Fig. 13 in \citet{2014AJ....148..122G}) shows exactly the same 
features as ours.

In addition, we collected 12 epochs of WISE data from the WISE All-Sky Single Epoch 
database. This includes photometry in the filter W1 and W2 (3.4$\,\mu m$ and 4.6$\,\mu m$), which are equivalent to the
Spitzer channels 1 and 2, but also W3 and W4 magnitudes at 12 and 22$\,\mu m$, a wavelength domain not covered anymore by
Spitzer. ISO-Oph-50 is only detected at s/n $>10$ in 8 out of 12 epochs in W3 and 5 out of 12 in W4, in the following we only
use these robust detections. Typical uncertainties of the WISE photometry are $<0.05$ for W1 and W2, $<0.1$ for W3 and W4. 
Compared with these uncertainties, the object is significantly variable in all four WISE channels, with amplitudes from maximum 
to mininum of 0.27, 0.17, 0.47, 0.7\,mag in W1-W4. The WISE colour $\mathrm{W1} - \mathrm{W2}$ increases by 0.2\,mag as the 
object gets fainter in W1.

The full mid-infrared lightcurve in the two short-wavelength bands is shown in Fig. \ref{irac}, which combines Spitzer
and WISE data. We also list the photometry in the appendix in tables \ref{at2} to \ref{at4}.

\begin{figure*}
\includegraphics[width=8.6cm]{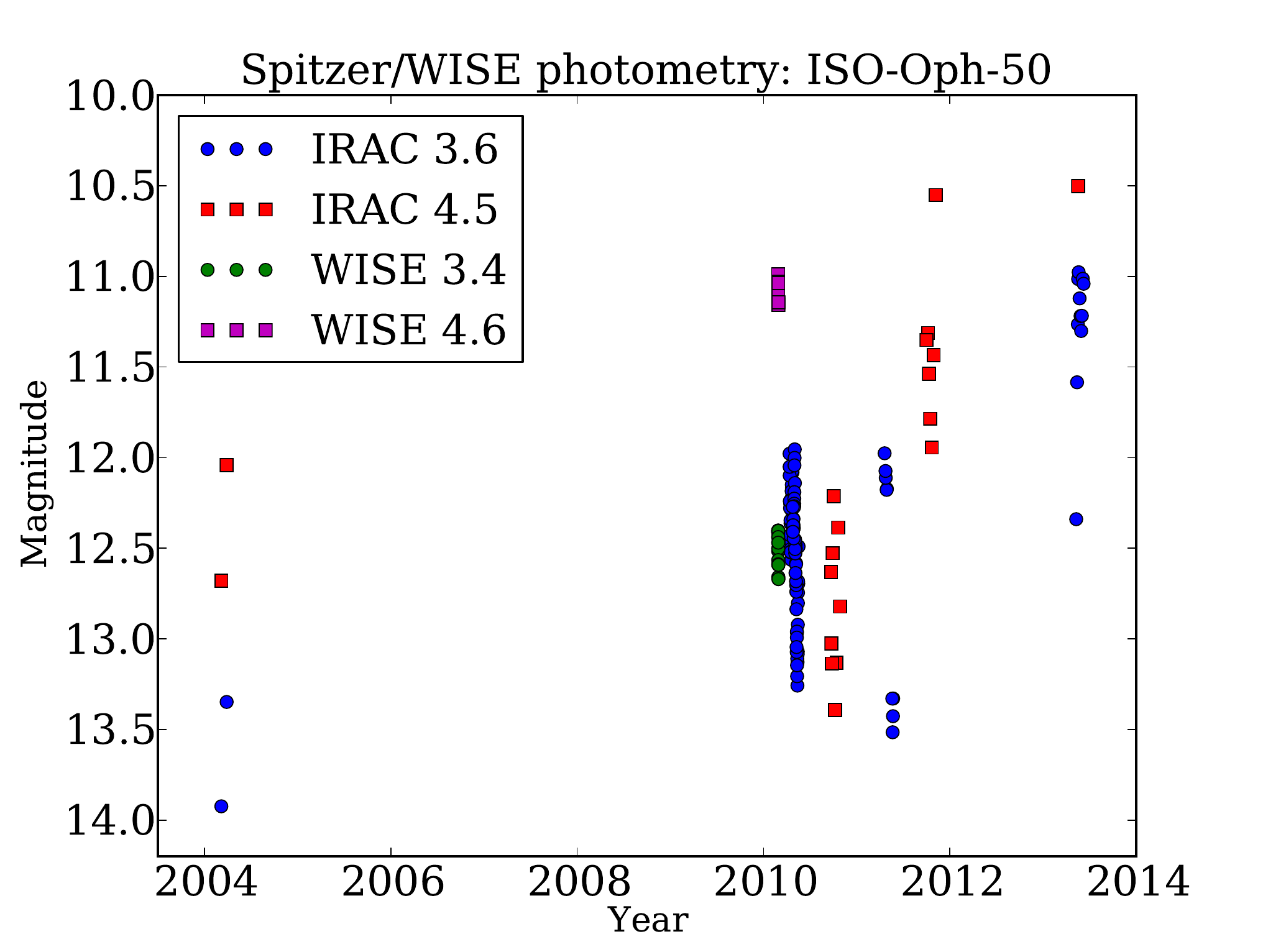}
\includegraphics[width=8.6cm]{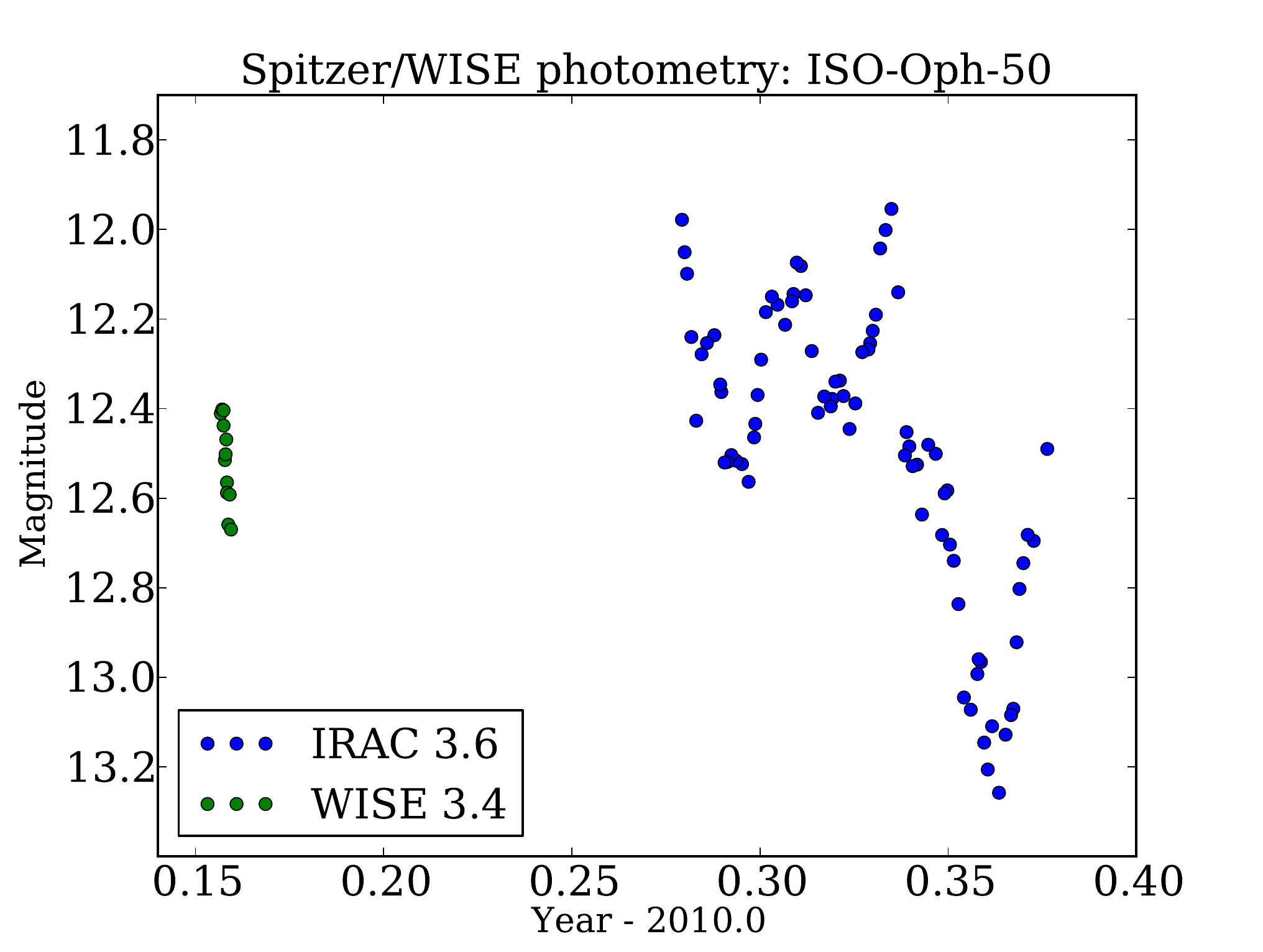}
\caption{\label{irac} Mid-infrared photometry from Spitzer and WISE for ISO-Oph-50. The left panel shows the full lightcurve,
the right panel a detailed view on the data from the year 2010 in the 3.4-3.6$\,\mu m$ band.}
\end{figure*}

\section{The complex variability of ISO-Oph-50}
\label{s2}

ISO-Oph-50 has now been observed for more than 20 years. The observational record prior to this paper has been summarised by 
\citet{2012A&A...539A.151A}, see in particular their Fig. 5, but a convincing interpretation of the data is still missing. 
Our continuous monitoring in the near-infrared, our spectroscopy, plus the new mid-infrared data now adds significantly to
the empirical constraints. In the following we will present the observational evidence in its entirety.

\subsection{Near-infrared lightcurve}
\label{brightfaint}

\begin{figure}
\includegraphics[width=9.0cm]{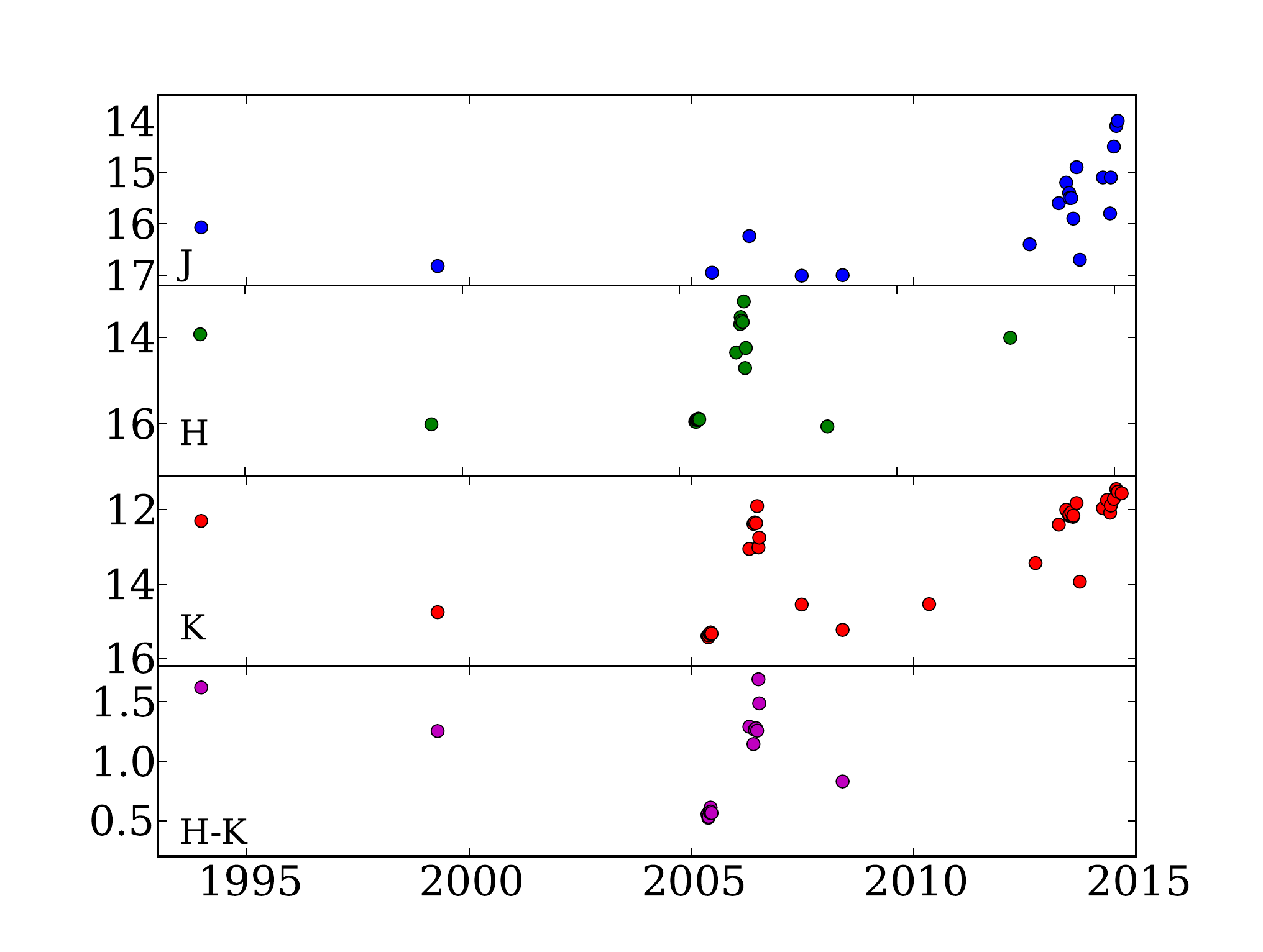}
\caption{\label{longlc} Long-term lightcurve for ISO-Oph-50, including data from various sources in the literature and our SMARTS
photometry. See Sect. \ref{brightfaint} for more details.}
\end{figure}

In Fig. \ref{longlc} we plot the long-term lightcurve for ISO-Oph-50 in the near-infrared bands and colours, combining all literature
data and our own new photometry. This is an updated version of Fig. 5 in \citet{2012A&A...539A.151A}, containing datapoints from 2MASS
\citep{2006AJ....131.1163S}, UKIDSS \citep{2007MNRAS.379.1599L} and a variety of other sources 
\citep{1997ApJS..112..109B,2008A&A...485..155A,2011ApJ...726...23G,2012ApJ...751...22B,2013MNRAS.430.2910S}.\footnote{In addition, the
object was not detected by DENIS in June 1999, which roughly implies $J>16$ and $K>14$. We note that the object is not covered by the 
second release of the VISTA Hemisphere Survey; later releases of the survey are expected to add another epoch.} The photometric records 
shows that the object is strongly variable. Its magnitude changes roughly from 12 to 16 in the K-band, from 13 to 16 in the H-band, and 
from 14 to 17 in the J-band.

There is some evidence for a bimodal behaviour, in the sense that the object spends prolonged periods of time near its
brightest and faintest state. Most of our own SMARTS datapoints from 2013-14 are near the brightest state around $K\sim 12$. The object
was similarly bright in 2006 in the lightcurve presented by \citet{2008A&A...485..155A}, with $K\sim 12-13$ and $H\sim 14$. An 
individual datapoint from 1993 is also at $K\sim 12$. On the other hand, the object was found to be significantly fainter in 2005
with $K\sim 15$ and $H\sim 16$. Datapoints from 1999 and 2007 are on a similar level. The near-infrared lightcurves show that
the transition from faint to bright and vice versa has to occur on a timescale significantly shorter than one year. The large
drop at the end of 2013 may indicate that the transition is as short as one month. In addition, Fig. \ref{longlc} may
indicate a gradual brightening between 2009 and 2014. Considering the years with photometry, the object seems to spend about half
of the time in bright state and the other half in faint state.

The large-scale variability evident in the long-term lightcurve is associated with an intriguing colour trend, first noted
by \citet{2008A&A...485..155A}. The object is usually significantly redder when bright, or conversely, bluer when faint. For 
example, between 2005 and 2006 ISO-Oph-50 became brighter by several magnitudes and redder by 0.8\,mag in $H-K$. Also, in 
2013-14 the object was bright ($K\sim 12$) and red ($J-K =3-4$), while in 2005, 2007 and 2008 it was faint ($K\sim 15$) and 
significantly bluer ($J-K\sim 2$). This behaviour, however, does not hold for all datapoints (see for example the 2MASS datapoints 
in 1999 when the object was faint and red). The colour trend is further discussed in Sect. \ref{s3}.

\begin{figure}
\includegraphics[width=9.0cm]{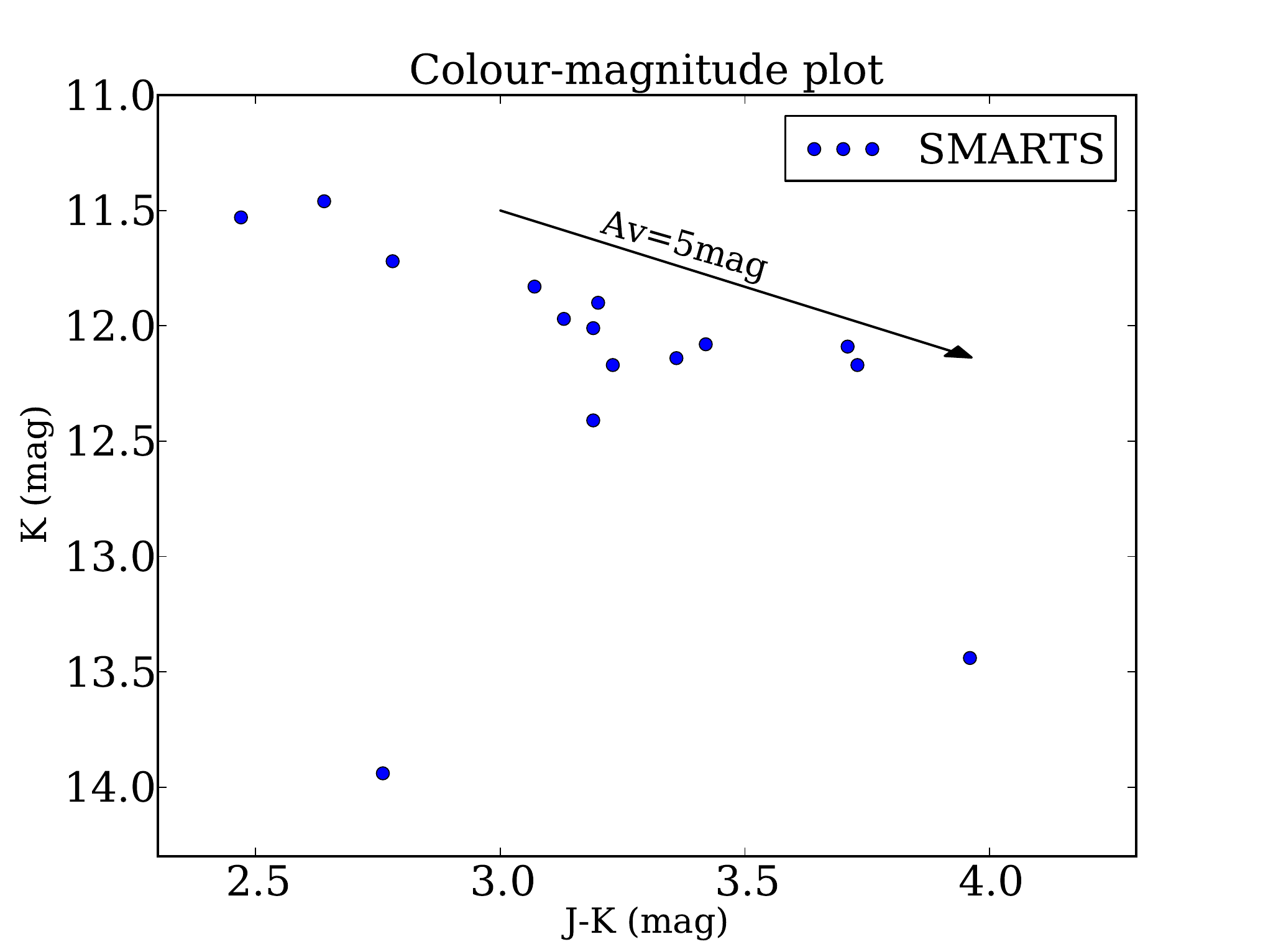}
\caption{\label{col} Colour-magnitude plot for the new SMARTS near-infrared photometry of ISO-Oph-50 covering the
years 2012-2014. An extinction vector is overplotted, using the reddening law by \citet{1990ARA&A..28...37M}.}
\end{figure}

Near its maximum flux, ISO-Oph-50 shows some additional variability on timescales of weeks with amplitudes less than 1\,mag in K-band, 
as seen in the 2006 lightcurve from \citet{2008A&A...485..155A} and in our SMARTS photometry from 2012-14 (see Fig. \ref{lc}). Our 
colour-magnitude plot (Fig. \ref{col}) demonstrates that during that period the object mostly becomes redder when fainter, i.e. a
colour trend opposite the one seen in the long-term evolution. The colour variability is following roughly
the extinction vector, which is overplotted in Fig. \ref{col}. This implies that variable extinction is likely the main driver 
for the bright state fluctuations observed for ISO-Oph-50. In the 2013-14 data, the $J-K$ colour varies by 1.4\,mag, which 
corresponds to optical $A_V$ variations by about 7\,mag. Again there are two interesting exceptions in 2012-14 which do not fit the
reddening vector. Compared with the reddening trend in the remaining (J-K,J) dataset, these two measurements are much too blue
for the K-band brightness. This could be the same 'blue when faint' effect seen in the long-term lightcurve (see above).

\subsection{Mid-infrared lightcurve}
\label{mir}

The mid-infrared photometry from Spitzer and WISE (Fig. \ref{irac}) shows evidence for variations by up to 3\,mag at 3.6$\mu m$ and 2\,mag
at 4.5$\,\mu m$, and also for variability at 12 and 22$\,\mu m$ (albeit with sparse sampling). This was already pointed out in 
\citet{2013MNRAS.430.2910S}. Overall, the object gradually became brighter from 2010 to 2013, which is similar to the trend seen in 
the near-infrared lightcurve over this timespan. 

In the year 2010 the object was frequently visited by WISE and Spitzer at 3.6$\,\mu m$, resulting in the only high-cadence lightcurve
for this object thus far (see Fig. \ref{irac}, right panel). The Spitzer dataset shows up to one magnitude of variations with 
a typical timescale of 1-2 weeks. These changes can be called 'quasi-periodic', in the sense that the brightness goes up and down
with some regularity, but both the amplitude and the period changes. This morphology could correspond to a series of eclipses of varying depths, 
with a decrease in flux by 0.2-0.8\,mag followed by an increase, without flat bottom. The lack of colour information makes it difficult
to constrain the nature of these variations any further. We note that \citet{2014AJ....148..122G} find a 'characteristic timescale' 
of 5.6\,d from an autocorrelation analysis of the 3.6$\,\mu m$ lightcurve of ISO-Oph-50, confirming that the dominant timescale of
this quasi-periodic behaviour is around one week.

\subsection{Near-infrared spectroscopy}

Our near-infrared spectra (Fig. \ref{spec}) show the typical spectral signature of a young mid M dwarf, 
with the broad peaks in H- and K-band caused by water absorption as well as the CO absorption bandheads at $>2.3\,\mu m$ 
\citep[e.g.][]{2009ApJ...702..805S}. The K-band shows lines at 2.21, and 2.26$\,\mu m$ (see Fig. \ref{spechk}), 
which are most likely caused by Na\,I and Ca\,I absorption \citep{2005ApJ...623.1115C}. Both lines
are typical for M3-M5 dwarfs. A few more possible photospheric lines are seen in the H-band (Fe\,H at 1.62, Al\,I at 1.67, and 
Mg\,I at 1.71$\,\mu m$); all three are again characteristic for the M1-M5 regime.

Based on the K-band photometry in the acquisition images, the object was near its brightest state when the spectra were
taken. As pointed out in Sect. \ref{brightfaint}, the near-infrared emission in bright state is subject to significant reddening.
To deredden the spectrum, we adopt an extinction of $A_v \sim 10$, which was estimated from the K-band magnitude 
of $\sim 12$ in the acquisition image and the colour trend seen in Fig. \ref{col}, assuming a photospheric $J-K$ of 1.0, which is 
appropriate for such an M-type object. This was 
done following the typical extinction vector shown in Fig. \ref{col}, using the reddening law by \citet{1990ARA&A..28...37M}. 
After dereddening the spectra, we estimate spectral types using two different indices and obtain M6 \citep[H2O index;][]{2007ApJ...657..511A} 
and M4 \citep[H2O-K2 index;][]{2012ApJ...748...93R}. As shown recently by \citet{2014MNRAS.442.1586D}, these two indices yield consistent
results within one subtype for young mid to late M-type objects, but while the H2O index is only calibrated for spectral types of M5 or later,
the H2O-K2 index covers the full M-type regime. Thus, we put more credence into the result from H2O-K2 and conclude that the 
spectral type is M4 with an uncertainty of one subtype.
For comparison, the only previous spectrum available was taken in the optical by \citet{2012A&A...539A.151A}. They estimate
an extinction of $A_V = 4.7$\,mag and a spectral type of M3.5, consistent with our value.

Our spectra show some evidence for ongoing accretion, based on the presence of Paschen $\beta$ and Bracket $\gamma$ emission 
features \citep{2004A&A...424..603N}. This further confirms the youth of the object. Interestingly, the optical spectrum in faint 
state does not show any H$\alpha$ emission or other evidence for accretion \citep{2012A&A...539A.151A}. 

\begin{figure}
\includegraphics[width=8.6cm]{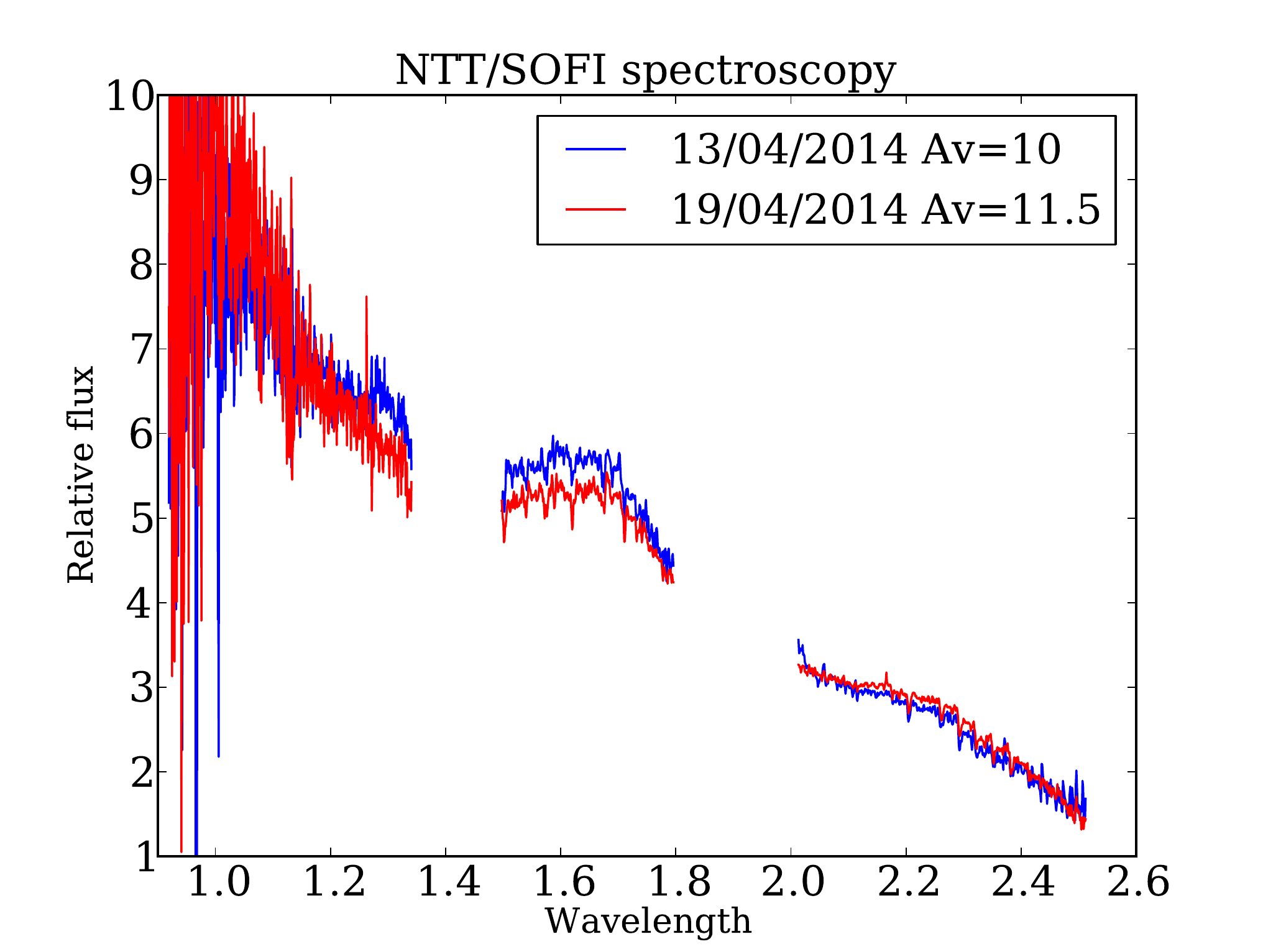}
\caption{\label{specdered} Dereddened spectra for ISO-Oph-50 taken on April 13th and 19th 2014.}
\end{figure}

Our two spectra, taken 6 days apart near lightcurve maximum ($K = 11.85$ and $12.0$), 
show significant spectral variability (Fig. \ref{spec}). The spectrum taken on April 13th 2014 is brighter and bluer than the one from
April 19th. This variation can approximately be accounted for by variable extinction. Assuming $A_V = 10$ for April 13th and
$A_V = 11.5$ for April 19th would result in quite similar dereddened spectra with remaining differences in the range of 10\%
(see Fig. \ref{specdered}). As pointed out in Sect. \ref{brightfaint} such extinction variations are consistent with the photometric 
lightcurve in the bright state.

\section{The central object: an underluminous very low mass star}
\label{central}

With the available spectroscopy and photometry we can place constraints on the nature of the central object. The estimated
spectral type of M4 would correspond to an effective temperature of about 3400$\pm 200$\,K \citep{2014ApJ...785..159M} and a mass of around 
0.4$\,M_{\odot}$, assuming an age of 1\,Myr \citep[BCAH isochrone;][]{1998A&A...337..403B}. Given the uncertainty in spectral 
typing and conversion to temperature, the plausible range for the mass is 0.2-0.6$\,M_{\odot}$. Thus, the central object is a
low mass star. Given the evidence for variability, accretion, strong extinction and excess emission from the disk, plus the 
uncertainty in the age and the evolutionary tracks, a more precise estimate of the mass is difficult to obtain.

According to the 1\,Myr BCAH isochrones, the K-band brightness for a 0.2-0.6$\,M_{\odot}$ star at the distance of $\rho$\,Oph (140\,pc) and
at an age of 1\,Myr would be 8 to 9.5\,mag. This is 5-7\,mag brighter than the minimum light of ISO-Oph-50; as already pointed 
out by \citet{2012A&A...539A.151A} the object is significantly underluminous. Thus, during the faint state most of the light
from the central source has to be blocked by optically thick material along the line of sight. A plausible explanation is that the plane
of the disk is approximately in the line of sight, i.e. we see the object through an edge-on disk. The colours near minimum are only 
slightly redder than expected for a mid M dwarf (according to BCAH, $H-K \sim 0.2-0.3$, $J-K\sim 1.0$), in line with the presence
of an edge-on disk, as in this case the radiation near minimum is expected to be dominated by scattered light, which makes the object 
appear bluer (for reference, see the models by \citep{1993ApJ...414..773K}). This is not unlike the
color-magnitude trends seen in other young sources seen through an edge-on disk \citep{2010MNRAS.409.1557S,2002ApJ...571L..51J}.

Even near maximum the object appears underluminous by 2-4\,mag in K and J, respectively. Extrapolating from the colour trend seen in Fig. 
\ref{col} to a photospheric $J-K=1.0$ results in a dereddened K-band brightness of $K\sim 10.5$, much closer to the expected 
value but still underluminous. This could mean that the source is only partially visible in bright state or that is in fact at the 
low-mass end of the plausible range of masses.

\section{The origin of the variability}
\label{s3}

\subsection{Summary of observational evidence}
\label{sumobs}

According to the analysis in the previous section, ISO-Oph-50 shows the following characteristics:

\begin{itemize}
\item{The central source is a very low mass star affected by strong reddening and excess infrared emission.}
\item{The object is underluminous at all times, but particularly so when near lightcurve minimum, by up to 
5-7\,mag in the K-band. This is indicative of an object seen through an edge-on disk.}
\item{The near-infrared lightcurve shows large-scale variations over timescales of months and years. During these variations
the object mostly becomes bluer when fainter.}
\item{Furthermore, the object exhibits small-scale changes in the near-infrared fluxes on timescales of weeks to months. 
Here the object becomes redder when fainter, consistent with changes in exctinction $A_V$ by several magnitudes.}
\item{In the mid-infrared the object show variability characteristics similar to the near-infrarad. High-cadence
monitoring reveals quasi-periodic changes on timescales of 1-2 weeks.}
\item{There is evidence for accretion when the object is in maximum light, but not in minimum.}
\end{itemize}
In the following we will investigate possible scenarios that may explain these properties. We emphasise that what we 
propose here is mostly a qualitative scenario for the main features of the source; detailed modeling is required to 
clarify details. 

\subsection{A clumpy disk seen at high inclination}

The most conservative statement about the cause of the variability is that it arises in the inner parts of the disk. 
This is supported by three pieces of evidence: a) Near- and mid-infrared lightcurve show the same trend. b) Short-term 
colour variations are consistent with changes in extinction along the line of sight; the timescales of weeks to months 
indicate that the material responsible for these changes has to be close, but not on the central object. c) Over long 
timescales the objects becomes redder when brighter; the opposite would be expected for any source of variability at 
the photospheric level (i.e. hot or cool spots). As pointed out in Sect. \ref{central}, the underluminosity as well 
as the colour near minimum can be understood by a disk seen at high inclination (i.e. close to edge-on). 

We suggest therefore that the most likely explanation for the complex variability of ISO-Oph-50 is a very low mass
star surrounded by an inhomogenuous ('clumpy') edge-on disk. Near its brightness maximum, the central source is directly 
visible, but seen through substantial and variable extinction of $A_V = 10-15$\,mag. The light from the star may 
have to travel through optically thin parts of the disk, either the upper layers or a gap in the midplane. In addition, 
the inner disk is visible to the observer, i.e. the source also appears bright in the mid-infrared. In minimum, the 
direct light from the central object, the thermal emission from the inner disk, and the emission from the
accretion flow as well are blocked by optically thick material along the line of sight. Instead, the object is seen 
in scattered light and thus appears bluer than in bright state. The change from bright to faint could be caused 
by a clump of material moving into the line of sight due to the disk rotation or due to precession of the disk. 
This scenario can qualitatively explain all the features listed in Sect. \ref{sumobs}.

If the inhomogeneities in the disk of ISO-Oph-50 are persistent features in Keplerian rotation, the timescale of the variations gives 
us some idea about the location of the features relative to the central source. The quasi-periodic changes seen at 3.6$\,\mu m$ in 2010 
occur with a typical cycle of 1-2 weeks. The small-scale variations seen in 2012-14, mostly consistent with variable extinction, also occur on 
timescales of a few weeks. This would correspond to a feature at a distance of 0.05 to 0.15\,AU in Keplerian rotation. For a low-mass 
star, this distance is well beyond the radius of magnetospheric truncation and just beyond the dust sublimation radius, i.e. the 
features would be placed close to the inner edge of the dusty disk. 

However, if the variability is caused by only one feature, we would expect to see a repeatable pattern, 
stable amplitudes or perhaps a flat portion of the lightcurve, and not the irregular variability 
discussed in Sect. \ref{s2}. More likely, a multitude of evolving features in the inner disk of different size and density is 
responsible for the complex variability. Some of them may be optically thick and thus obscure the central source when passing 
through the line of sight. Others are optically thin and only affect the light through variable extinction. With multiple 
features, the distance from the central object could be significantly larger than the value derived above for the single-occulter 
scenario. The fact that ISO-Oph-50 is variable all the time tells us that the inhomogeneities are present in a major portion of the 
inner disk. Furthermore, the lack of pattern indicates that some of the clumps may be transient features that are formed on timescales 
of weeks to years and decay on similar timescales. 

The large amplitude of the variations, the lack of a clear long-term period, the seemingly erratic behaviour, the various irregularities, 
and the colour trend observed in the long-term variations distinguish this object from other young low-mass stars with evidence for 
obscuration by disk material, for example KH15D \citep{2001ApJ...554L.201H}, AA Tau \citep{2003A&A...409..169B,2013A&A...557A..77B}, or 
WL4 \citep{2008ApJ...684L..37P}. \citet{2013AJ....145...66F} find that a majority of stars with disks shows mid-infrared variations,
albeit with much smaller amplitudes than ISO-Oph-50, concluding that rapid changes in the inner disk are quite common. 
Based on near-infrared lightcurves, \citet{2001AJ....121.3160C} and, more recently, \citet{2013ApJ...773..145W} also find objects
with variability indicating changes in the inner disk. Some of the objects discussed by \citet{2013ApJ...773..145W} show
puzzling changes in the colour trend, similar to ISO-Oph-50 (see for example RWA\,2 and RWA\,7 in their
paper).

The behaviour of ISO-Oph-50 is in some ways reminiscent of UX Ori-type Herbig Ae/Be stars 
\citep[e.g.][]{1991Ap&SS.186..283G,2000A&A...364..633N,2003ApJ...594L..47D}. These objects are usually described as
stars with nearly edge-on disks in which clumps of dust and gas are temporarily in the line of sight and cause irregular,
long eclipses. Since the obscuration events in UX Ori stars occur on timescales of weeks to months, the occulting clumps have to be 
located in the puffed-up inner rim of the disk, rather than the outer flaring disk \citep[][see their Fig. 1]{2003ApJ...594L..47D}. 
ISO-Oph-50 could be a version of the same phenomenon, but with a central object that is by an order of magnitude less massive. For 
such low-luminosity objects, the scale of the disk is different. The inner edge of the dusty disk is closer to the central object 
($<<0.1$\,AU). Flaring dominates the disk geometry between 0.1 and 
1\,AU \citep[see the modeling in][]{2007ApJ...660.1517S,2009MNRAS.398..873S}. Thus, in contrast to its more massive counterparts, 
the occulters in the disk of ISO-Oph-50 are not necessarily located at the inner edge of the disk, they could be spread over a 
range of distances. 

Thus, compared with the literature, ISO-Oph-50 is an extreme case of YSO variability, but following trends which have been observed
on a smaller scale in other young sources. The physical origin of the inhomogeneities is difficult to constrain from the existing data. 
The clumps could be density enhancements (e.g, spiral structure or vortices) or geometrical anomalies (e.g., warps or walls), as found 
for other young stellar objects. It is possible that the extreme behaviour of ISO-Oph-50 is solely due to a high inclination of the disk.
Other explanations include a recent perturbation of the disk, for example by an accretion burst or by an instability.

\subsection{Alternative scenarios}

Here we briefly consider two possible alternative scenarios proposed in the literature for the behaviour of ISO-Oph-50. Based on the flux 
increase between the 2005 and 2006 lightcurve, \citet{2008A&A...485..155A} conclude that ISO-Oph-50 could 
be an object undergoing an accretion burst. Since it was observed to be bright before, this would be a recurrent phenomenon, 
perhaps comparable to the prototype EX Lupi. Including the more recent data, there are several arguments against this scenario. 
First, as already pointed out in \citet{2013MNRAS.430.2910S}, EX Lupi is observed in outburst only about 10\% of the time 
\citep{2007AJ....133.2679H}, whereas ISO-Oph-50 could be near maximum about half of the time (see Sect. \ref{brightfaint}). 
Second, the source is bluer when fainter, which is not what is observed in EX Lupi-type objects \citep{2009AJ....138.1137A}.
Third, ISO-Oph-50 is clearly too faint for its spectral type, even in its bright state (Sect. \ref{central}), which essentially 
excludes that the faint state represents the quiescent period between two accretion bursts. And fourth, the object does not bear
the typical spectroscopic signatures of an FU Ori burst (high veiling, no photospheric absorption lines, excess CO absorption 
and emission, see \citet{2010AJ....140.1214C}). For all these reasons, accretion bursts alone do not provide a satisfactory 
explanation for the observed variability, but could contribute to some of the variations.

\citet{2012A&A...539A.151A} speculate that ISO-Oph-50 consists of 'two objects with similar spectral type, one having strong 
IR excess and undergoing variable circumstellar extinction by the disk (i.e., close to edge-on), the other component either disk free 
or at least not obscured by its disk.' At maximum brightness, both stars would be visible. In minimum, the primary would be obscured 
by its disk, the light thus be dominated by the secondary. Since the secondary is disk-less, the emission would become bluer.
As outlined above, changes in circumstellar extinction do explain parts of the variability, but there is no independent 
evidence for ISO-Oph-50 being a binary. If it is a binary, we would expect the object to be overluminous near maximum when both
components are visible, whereas in reality it may still be fainter than the intrinsic brightness of one star with the appropriate 
spectral type (see Sect. \ref{central}). The assumption that the system contains a disk-less companions adds complications to the
proposed scenario, but cannot be excluded based on the available data.

\subsection{Further observations}

Continued and more detailed observations of ISO-Oph-50 could provide a wealth of information about the spatial distribution of 
dust in the inner disk. The structures discussed here are too small for imaging with high spatial resolution, and the object too 
faint for optical/infrared interferometry. The dominance of scattered light in faint state could be verified with polarimetric 
observations, but that as well will be challenging for this faint source. If the proposed scenario outlined above is correct, one 
might also expect significant changes in spectroscopic features in the mid-infrared (e.g., ice feature, see \citet{2005ApJ...622..463P}). 
Apart from these specific tests, the best strategy to map the circumstellar environment of ISO-Oph-50 is prolonged multi-band 
(or spectrophotometric) monitoring in the near/mid-infrared. To test the structure of the outer disk, complementary submm/mm 
interferometry would be beneficial. Since the object is currently near its brightest state, it currently presents us with a 
unique chance to obtain more detailed insights.

\section*{Acknowledgments}
We thank Catarina Alves de Oliveira and Ray Jayawardhana for useful comments on an early draft of this paper. 
This publication makes use of images obtained with SMARTS telescopes; AS is grateful for the support received through 
the collaboration. This work makes use of observations made with the Spitzer Space Telescope, which is operated by the Jet 
Propulsion Laboratory, California Institute of Technology under a contract with NASA. Part of this work was funded by the 
Science Foundation Ireland through grant no. 10/RFP/AST2780. 

\appendix

\section{Photometry tables}
\label{a1}

\begin{table}
\caption{SMARTS photometry for ISO-Oph-50. Typical errors are 0.3\,mag in I, 0.1\,mag in J, and 0.05\,mag in H and K.
\label{at1}}
\begin{tabular}{lllll}
\hline
Epoch  & I    & J    & H    & K  \\
date   & mag  & mag  & mag  & mag \\
\hline
120812 & 20.1  & 16.4 & 14.1 & --\\
120925 & 19.8  & 17.4 & --   & 13.44 \\
130404 & 19.1  & 15.6 & --   & 12.41 \\
130614 & --    & 15.2 & --   & 12.01 \\
130628 & 19.7  & 15.4 & --   & 12.17 \\
130701 & 19.6  & 15.5 & --   & 12.14 \\
130716 & 19.6  & 15.5 & --   & 12.08 \\
130730 & $>19$ & --   & --   & 12.20 \\
130801 & --    & 15.9 & --   & 12.17 \\
130828 & --    & 14.9 & --   & 11.83 \\
130924 & $>20$ & 16.7 & --   & 13.94 \\
140401 & 19.5  & 15.1 & --   & 11.97 \\
140425 & 19.4  & --   & --   & --    \\
140505 & 19.0  & --   & --   & 11.75 \\
140530 & 19.3  & 15.8 & --   & 12.09 \\
140605 & 18.3  & 15.1 & --   & 11.90 \\
140630 & 18.4  & 14.5 & --   & 11.72 \\
140720 & 17.6  & 14.1 & --   & 11.46 \\
140801 & 17.7  & 14.0 & --   & 11.53 \\
140902 & 17.9  & --   & --   & 11.57 \\
\hline
\end{tabular}
\end{table}

\begin{table*}
\caption{WISE multi-epoch photometry for ISO-Oph-50. Only values with signal-to-noise ratio $>10$ listed.
\label{at2}}
\begin{tabular}{lllllllll}
\hline
MJD         & W1    & W1err  & W2      & W2err  & W3     & W3err  & W4  & W4err \\
d             & mag   & mag    & mag     & mag    & mag    & mag    & mag & mag   \\
\hline
55253.7066  & 12.411 & 0.030 & 11.003  & 0.025  & 8.120  & 0.053  & --     & --    \\
55253.8389  & 12.402 & 0.029 & 10.995  & 0.028  & 8.179  & 0.046  & --     & --    \\
55253.9714  & 12.404 & 0.033 & 11.025  & 0.031  & --     &  --    & --     & --    \\
55253.9712  & 12.438 & 0.032 & 10.989  & 0.026  & 8.406  & 0.082  & 4.543  & 0.073 \\
55254.1037  & 12.515 & 0.031 & 11.029  & 0.031  & 8.099  & 0.044  & 5.201  & 0.065 \\
55254.1697  & 12.502 & 0.032 & 11.107  & 0.033  & 8.069  & 0.044  & 5.051  & 0.055 \\
55254.2360  & 12.469 & 0.034 & 11.069  & 0.036  & 7.973  & 0.043  & --     & --    \\
55254.3021  & 12.565 & 0.036 & 11.104  & 0.032  & --     &    --  & 4.943  & 0.056 \\
55254.3022  & 12.588 & 0.038 & 11.037  & 0.029  & --     &    --  & 3.509  & 0.078 \\
55254.4345  & 12.659 & 0.034 & 11.144  & 0.030  & 7.937  & 0.036  & --     & --    \\
55254.5668  & 12.592 & 0.035 & 11.157  & 0.030  & 8.127  & 0.051  & --     & --    \\
55254.6991  & 12.670 & 0.033 & 11.143  & 0.033  & --     & 0.493  & --     &  --   \\
\hline
\end{tabular}
\end{table*}

\begin{table}
\caption{Spitzer/IRAC 3.6$\,\mu$ photometry for ISO-Oph-50. Typical errors are 0.05\,mag.
\label{at3}}
\begin{tabular}{llll}
\hline
MJD  & 3.6$\,\mu m$ & MJD & 3.6$\,\mu m$\\
d    & mag          & d   & mag \\
\hline
53071.3221 & 13.92  & 55320.5253 & 12.48 \\
53092.2178 & 13.35  & 55320.8490 & 12.53 \\
55298.4622 & 11.98  & 55321.2717 & 12.53 \\
55298.7187 & 12.05  & 55321.7550 & 12.64 \\
55298.9558 & 12.10  & 55322.3612 & 12.48 \\
55299.3698 & 12.24  & 55323.0907 & 12.50 \\
55299.8427 & 12.43  & 55323.7057 & 12.68 \\
55300.3675 & 12.28  & 55323.9537 & 12.59 \\
55300.8875 & 12.25  & 55324.2156 & 12.58 \\
55301.6086 & 12.24  & 55324.4720 & 12.70 \\
55302.1721 & 12.35  & 55324.8390 & 12.74 \\
55302.2817 & 12.36  & 55325.2917 & 12.84 \\
55302.5990 & 12.52  & 55325.8245 & 13.05 \\
55302.8846 & 12.52  & 55326.4895 & 13.07 \\
55303.2533 & 12.50  & 55327.1309 & 12.99 \\
55303.7152 & 12.52  & 55327.2521 & 12.96 \\
55304.2878 & 12.52  & 55327.4740 & 12.97 \\
55304.9309 & 12.56  & 55327.7916 & 13.15 \\
55305.4586 & 12.46  & 55328.1405 & 13.21 \\
55305.5766 & 12.43  & 55328.5600 & 13.11 \\
55305.8023 & 12.37  & 55329.2289 & 13.26 \\
55306.1441 & 12.29  & 55329.8696 & 13.13 \\
55306.6025 & 12.18  & 55330.4020 & 13.08 \\
55307.1870 & 12.15  & 55330.6184 & 13.07 \\
55307.7318 & 12.17  & 55330.9363 & 12.92 \\
55308.4711 & 12.21  & 55331.2188 & 12.80 \\
55309.1426 & 12.16  & 55331.5915 & 12.75 \\
55309.2837 & 12.14  & 55332.0153 & 12.68 \\
55309.6008 & 12.07  & 55332.5999 & 12.70 \\
55310.0027 & 12.08  & 55333.3515 & 16.74 \\
55310.4749 & 12.15  & 55333.9147 & 12.49 \\
55311.0558 & 12.27  & 55671.0198 & 11.98 \\
55311.6647 & 12.41  & 55674.5251 & 12.07 \\
55312.2659 & 12.37  & 55675.5156 & 12.11 \\
55312.9024 & 12.40  & 55679.0441 & 12.18 \\
55313.0397 & 12.38  & 55679.9347 & 12.17 \\
55313.3508 & 12.34  & 55701.0738 & 13.33 \\
55313.7856 & 12.34  & 55702.4730 & 13.52 \\
55314.1361 & 12.37  & 55703.7944 & 13.43 \\
55314.7244 & 12.45  & 55704.6386 & 13.33 \\
55315.2896 & 12.39  & 56421.9529 & 12.34 \\
55315.9590 & 12.27  & 56425.3507 & 11.58 \\
55316.5588 & 12.27  & 56428.4775 & 11.26 \\
55316.7232 & 12.25  & 56429.5838 & 11.01 \\
55316.9729 & 12.23  & 56431.6701 & 10.98 \\
55317.2813 & 12.19  & 56435.3176 & 11.12 \\
55317.6984 & 12.04  & 56439.0334 & 11.22 \\
55318.2266 & 12.00  & 56441.4309 & 11.30 \\
55318.7913 & 11.95  & 56444.1133 & 11.22 \\
55319.4430 & 12.14  & 56447.7893 & 11.01 \\
55320.1009 & 12.50  & 56450.8949 & 11.04 \\
55320.2579 & 12.45  & 56795.8969 & 10.93 \\
\hline
\end{tabular}
\end{table}

\begin{table}
\caption{Spitzer/IRAC 4.5$\,\mu$ photometry for ISO-Oph-50. Typical errors are 0.05\,mag.
\label{at4}}
\begin{tabular}{ll}
\hline
MJD  & 4.5$\,\mu m$ \\
d    & mag          \\
\hline
53071.3221  & 12.68 \\
53092.2178  & 12.04 \\
55461.3747  & 12.63 \\
55462.4175  & 13.03 \\
55464.3508  & 13.14 \\
55467.3647  & 12.53 \\
55471.3766  & 12.21 \\
55476.8021  & 13.39 \\
55482.4433  & 13.13 \\
55489.2337  & 12.39 \\
55496.7011  & 12.82 \\
55835.2529  & 11.35 \\
55841.3555  & 11.31 \\
55845.2345  & 11.54 \\
55850.1141  & 11.79 \\
55856.1044  & 11.94 \\
55863.2050  & 11.44 \\
55871.3986  & 10.55 \\
56429.5838  & 10.50 \\
56795.8969  & 10.49 \\
\hline
\end{tabular}
\end{table}

\newcommand\aj{AJ} 
\newcommand\actaa{AcA} 
\newcommand\araa{ARA\&A} 
\newcommand\apj{ApJ} 
\newcommand\apjl{ApJ} 
\newcommand\apjs{ApJS} 
\newcommand\aap{A\&A} 
\newcommand\aapr{A\&A~Rev.} 
\newcommand\aaps{A\&AS} 
\newcommand\mnras{MNRAS} 
\newcommand\pasa{PASA} 
\newcommand\pasp{PASP} 
\newcommand\pasj{PASJ} 
\newcommand\solphys{Sol.~Phys.} 
\newcommand\nat{Nature} 
\newcommand\bain{Bulletin of the Astronomical Institutes of the Netherlands}
\newcommand\memsai{Mem. Soc. Astron. Ital.}
\newcommand\apss{Astrophysics and Space Science}

\bibliographystyle{mn2e}
\bibliography{aleksbib}

\label{lastpage}

\end{document}